\documentclass{article}

\usepackage[margin=1in]{geometry}

\usepackage{textcomp} 

\usepackage{graphicx}
\usepackage[labelformat=simple]{subfig}

\usepackage{amsmath}
\usepackage{xcolor}

\usepackage[T1]{fontenc}
\usepackage[utf8]{inputenc}              
\usepackage[italian]{babel} 

\newcommand{\caporali}[1]{«#1»}

\newcommand{\Boltz}{ k_{\rm\scriptscriptstyle B}}

\newcommand{\Tr}{{\rm Tr}}

\newcommand{\cu}{{\hat{\pmb{n}}}}

\title{The fourth law of thermodynamics: steepest entropy ascent}

\author{Gian Paolo Beretta\\
	\small Università di Brescia, Italy
}

\date{November 18, 2019\\To appear in Philosophical Transactions of the Royal Society A}

\begin{document}
	\maketitle

\begin{abstract}
When thermodynamics is understood as the science (or art) of constructing effective models of natural phenomena by choosing a minimal level of description capable of capturing the essential features of the physical reality of interest, the scientific community has identified a set of general rules that the model must incorporate if it aspires to be consistent with the body of known experimental evidence. Some of these rules are believed to be so general that we think of them as laws of Nature, such as the great conservation principles, whose "greatness" derives from their generality, as masterfully explained by Feynman in one of his legendary lectures. The second law of thermodynamics is universally contemplated among the great laws of Nature. In this paper we show that, in the past four decades, an enormous body of scientific research devoted to modeling the essential features of nonequilibrium natural phenomena has converged from many different directions and frameworks towards the general recognition (albeit still expressed in different but equivalent forms and language) that another rule is also indispensable and reveals another great law of Nature that we propose to call the \caporali{fourth law of thermodynamics}. We state it as follows:   every nonequilibrium  state of a system or local subsystem for which entropy is well-defined must be equipped with a metric in state space  with respect to which the irreversible component of its time evolution is in the direction of steepest entropy ascent compatible with the conservation constraints. To illustrate the power of the fourth law, we derive (nonlinear) extensions of Onsager reciprocity and  fluctuation-dissipation relations to the far-nonequilibrium realm  within the framework of the rate-controlled constrained-equilibrium (RCCE) approximation (also known as the quasi-equilibrium approximation).
\end{abstract}

\section{Introduction}

The first  and the second law of thermodynamics are considered among the ``great laws of Nature.'' What we mean by this is vividly explained by Feynman in one of his legendary lectures \cite{Feynman}: a ``great law of Nature'' is a rule, a feature, an assertion that the scientific community has grown to consider an indispensable element of any successful model of a natural phenomenon, at any level of description.
The main objective in this paper, is to point to a feature that has emerged from scientific progress in the past few decades and has become a key,  indispensable element of all successful models of nonequilibrium natural phenomena. For this reason, we claim that this feature has effectively grown to the level of a new great law of Nature, that we propose to call ``the fourth law of thermodynamics.''

To make the present discussion as precise as possible, we propose to adopt the following two distinct  meanings of the word \caporali{thermodynamics}:
(1) ``Applied Thermodynamics'' is the art of modeling the kinematics and the dynamics of physical systems by choosing the most appropriate level of description for the 'application of interest' and implementing/exploiting the general principles/rules/laws that any such model ought to satisfy to guarantee a fair representation of the physical reality it is meant to describe (in the sense of Margenau's ``plane of perceptions'' \cite{Margenau}). (2) ``Foundational Thermodynamics'' is the art of extracting/distilling/identifying such general principles/rules/laws from the successes and failures of the entire body of scientific modeling efforts to rationalize experimental observations.

So, when thermodynamics is understood as the science/art of constructing effective models of natural phenomena by choosing a minimal level of description capable of capturing the essential features of the physical reality of interest, the scientific community has identified a set of general rules that the model must incorporate if it aspires to be consistent with the body of known experimental evidence. Some of these rules are believed to be so general that we think of them as laws of Nature, such as the great conservation principles, whose ``greatness'' derives from their generality.

\begin{figure}[!ht]
	\centering		\includegraphics[width=\textwidth]{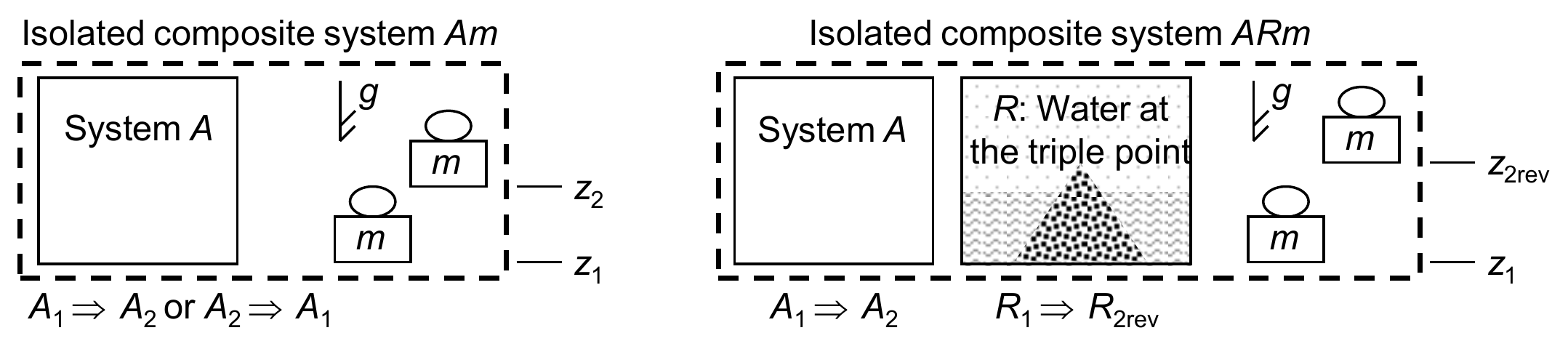}
	\caption{(Left) The first law  guarantees that 
		any pair of states $A_1$ and $A_2$ of a (well-separated) system $A$ (fixed volume $V$) can be the end states of a  process for the isolated composite $Am$, 
		where $m$ is a weight in a uniform gravity acceleration $g$. Measuring $(z_1-z_2)mg$ in such a process defines the  
		energy difference  $E^A_2-E^A_1$ for the two states of $A$. (Right) The second law guarantees that the same two states can be the end states of a reversible process for the isolated composite $ARm$, where $R$ is a container in which pure water remains at the triple point. Measuring $E^R_1-E^R_2$ in such a reversible process and dividing it by $273.16\ \rm  K$ defines the  entropy
		difference $S^A_2-S^A_1$ for the two states of $A$.}
	\label{Figure1}
\end{figure}

The \caporali{first law of thermodynamics} \cite[p.30]{Book} requires that --- regardless of the details of the model assumed to describe a ``physical system'' $A$ (any physical system) and its ``states''\footnote{By state we generally mean the collection of the values of all the properties at one instant of time, and by property a physical observable defined by a measurement procedure that produces an outcome that depends on a single instant of time.} --- for any two states $A_1$ and $A_2$ in which  $A$ is isolated and  uncorrelated from the rest of the universe,  it must be admissible within the model to devise at least one  time evolution in which $A_1$ and $A_2$ are the end states of the system, while the only effect in the rest of the universe is a change in elevation of a weight in a gravity field (or an equivalent work element  \cite[App.C]{HG2}). Moreover, for a given weight and gravity acceleration the change in elevation is the same for any such time evolution between states $A_1$ and $A_2$.  Such requirement is necessary to support the measurement procedure \cite[p.32]{Book}, illustrated in Figure \ref{Figure1}(left), that defines operationally the ``energy difference'' between any two states in which the system is isolated and  uncorrelated. In addition, it implies the additivity of energy differences for noninteracting composite systems, the conservation of energy and, therefore, the energy balance equation.

The \caporali{second law of thermodynamics} \cite[p.62]{Book} requires that --- again, regardless of the details of the model assumed to describe a physical system $A$ and its states --- for any two states $A_1$ and $A_2$ in which  $A$ is isolated and  uncorrelated from the rest of the universe,  it must be admissible within the model to devise at least one  reversible time evolution in which the system starts in state $A_1$ and ends in state $A_2$, while the only effects in the rest of the universe are a change in elevation of a weight in a gravity field and the change from state $R_1$ to state $R_2$ of a thermal reservoir (or heat bath) such as a container with water at the triple point in both states $R_1$ and $R_2$ (for more rigorous definitions see \cite{Book,ZBEntropy14,AAPP19}). By ``reversible'' we mean that the model must admit also a time evolution that returns system $A$ from state $A_2$ back to state $A_1$, while the only effects in the rest of the universe are the return of the weight to its original elevation and the change from state $R_2$ back to state $R_1$ of the thermal reservoir. 
Such requirement is necessary to support the measurement procedure \cite[p.102]{Book}, illustrated in Figure \ref{Figure1}(right), that defines operationally the ``entropy difference'' between any two states in which the system is isolated and  uncorrelated. In addition, it implies the additivity of entropy differences for composite systems in uncorrelated states, the conservation of entropy in reversible processes, the principle of non-decrease of entropy  and, therefore, the entropy balance equation.  

We emphasize that the present discussion focuses on when a given specific model has been chosen and set for the nonequilibrium problem of interest. This means that a given level and framework of description (e.g., macroscopic, mesoscopic, microscopic, classical, quantum, stochastic) has been chosen together with a specific set of state variables and a specific law for their time evolution, and that all definitions, including those of (local) energy, (local) entropy, and (ir-)reversibility, must be self-consistent within the assumed model. Therefore, the otherwise interesting discussions about how to define rules, such as coarse graining and projection methods, for passing consistently from a given level of description to a more macroscopic one \cite{Ottinger98,Grmela14,Pavelka14,Montefusco18}, about model reduction techniques \cite{Chiavazzo07,Lebiedz10,Pope13}, or about how to identify rate controlling constraints \cite{Rivadossi18}, do not play a role here.\footnote{However, from our claim in this paper, namely that the fourth law should apply within any level of description that contemplates dissipation, it follows that coarse graining, projection methods, and other rules to pass from one level to more macroscopic ones should also include the relations that must hold between the two steepest-entropy-ascent metrics that characterize the two related levels of description.}

The second law  is universally contemplated among the great laws of Nature, although no two scientists will tell you what it is in the same way, except when they agree to coauthor a paper (see, e.g.,  \cite{HG1,HG2,HG3,HG4,GBSecond15,LY99,Hatsopoulos08,LY13,LY14,ZBEntropy14,Brandao15,Weilenmann16,AAPP19}) or a book (see, e.g., \cite{HK,HGThermionic79, Book,OConnellHaile05,EbelingSokolov05,LebonJouCasas08,Kjelstrup08}). Our understanding of the laws of thermodynamics has never stopped evolving over the past two centuries. The initial focus on classical statistics and kinetic theory (Boltzmann), chemical kinetics and equilibrium (van't Hoff, Gibbs), quantum statistics (Fermi-Dirac, Bose-Einstein), near equilibrium and chemical kinetics (Onsager, Prigogine), shifted in more recent decades towards complex fluids and solids, far nonequilibrium, and small and quantum systems. On and off during this evolution, some of the basic concepts needed to be revisited to adapt/extend their applicability to the new realm of phenomena of interest. Questions like ``What is work?'', ``What is heat?'' \cite{GBHeat15,Weimer08,Levy12,Skrzypczyk14,Binder15,Esposito15,Sparaciari17}, ``What is entropy?'' \cite{HG1,HG2,HG3,HG4,Bender02,ZBInTech11,Santos11,Brandao13,Weilenmann16,YungerHalpern16}, ``What is macroscopic?'' \cite{GBSimple15,Goldstein15,Jarzynski17} have risen to a currently urgent need in the quantum (Q) communities (Q information, Q computing, Q thermal machines, Q fluctuations).\footnote{As already mentioned, the first law entails the existence of property energy for all states of every ``system'' by supporting its operational definition \cite[p.32]{Book} (see also \cite{Zanchini86,Zanchini88,Zanchini92}), but it can do so only for models in which the system is well separated from its environment. In the quantum framework this means that the effects of the environment on the system can be modeled via the dependence of the Hamiltonian operator on a set of classical control parameters. Suppose system $AB$ (Alice and Bob as a couple) is well separated but the influence of  Alice on Bob and viceversa is described by a full-fledged interaction Hamiltonian $V_{AB}$: then the energy of $AB$ is defined (represented by the mean value of the Hamiltonian $H_{A}+H_{B}+V_{AB}$) but the individual energies of $A$ and $B$ are not! The same issue is faced when $B$ (Bath) is the environment of $A$, hence, the difficulties in applying thermodynamic concepts to open systems unless the effects of the full-fledged system-bath interaction can be reduced to a description in terms of local operators such as in the Kraus--Kossakowski--Ingarden--Lindblad--Gorini--Sudarshan--Spohn (KKILGSS) models \cite{Kraus71,Kossakowski72,Ingarden75,Lindblad76,Gorini76,Spohn76} or in the locally-steepest-entropy-ascent (LSEA) models of composite systems \cite{Cimento2,Bregenz09,IJQT07,Cano15,Smith16}.}

The second law has been stated in many ways over the almost two centuries of  history of thermodynamics, and it is not our purpose here to review them. However, our preference goes to the Hatsopoulos-Keenan statement \cite[p.62]{Book} not only because we have provided rigorous proofs that it entails the better known traditional statements  (Kelvin-Planck \cite[p.64]{Book}, Clausius \cite[p.134]{Book}, Carath\'eodory \cite[p.121]{Book}), but --- quite importantly for the current and recent developments of nonequilibrium and  quantum thermodynamics --- because we have shown in \cite{ZBInTech11,ZBEntropy14,AAPP19} that the  operational definition of entropy supported by this statement is valid not only for the stable equilibrium states of macroscopic systems but also for their nonequilibrium states and it provides a solid  basis  for its extension to  systems  with only few particles and quantum systems.\footnote{Its extendability to correlated states of interacting or non-interacting systems is instead still the subject of intense debate, because the correlation entropy (often called mutual information), similarly to the mean energy of interaction between the subsystems, is a well-defined feature  for the overall state of the composite system, but there is no unique nor fundamental recipe to allocate it among the subsystems nor to assign it to the local (reduced, marginal) states of the subsystems, even though in the context of LSEA models we have  proposed a possible way in \cite[Eq.12]{Cimento2}, \cite[Sec.10]{Bregenz09}, and \cite[Eqs.60-61]{ROMP}.}  We have also shown that when restricted to macroscopic systems in equilibrium (in the sense of what we called ``simple system model of stable equilibrium states'' \cite[Ch.17]{Book})  our operational definition of entropy based on the Hatsopoulos-Keenan statement is essentially equivalent to the definition of entropy more recently proposed by Lieb and Yngvason in  \cite{LY99} and is closely related to the various extensions and improvements proposed thereafter \cite{LY13,LY14}.

Another important implication of the second law is the ``state principle,'' which asserts \cite[Ch.8]{Book}) that the equilibrium states of a system form an $(r+s+1)$--parameter family, where $r$ denotes the number of conserved properties in addition to energy and $s$ the number of control parameters of the Hamiltonian. This assertion, in turn, implies the  existence for every system of a concave ``fundamental equilibrium relation,'' for example,  $S=S_{\rm eq}(E,V,n_1,\dots,n_r)$,  for a system with volume $V$ as the only parameter and  $r$  different types of independently conserved constituents (amounts  denoted by $n_i$). The ``greatness'' of this second-law consequence stems from the fact the existence and concavity must hold for any system, but the functional dependence of the relation varies from system to system and is in fact what characterizes its equilibrium properties.

By analogy, and to allow full flexibility of formulation, what we propose to call the \caporali{fourth law of thermodynamics}  is any  assertion that --- regardless of the specific and technical details that are peculiar to one or the other non-equilibrium theory, or of the prose preferences of the different authors --- entails  a principle of existence of a metric field, defined over the entire state space of the modeled system,  with respect to which the irreversible (dissipative) component of the time evolution of the system (or of each of its subsystems) is steepest entropy ascent (SEA). The functional dependence of the SEA metric on the state variables varies from system to system and is in fact what characterizes its nonequilibrium behavior.

In Sections \ref{Sec2} and \ref{Sec3} we prepare the stage for the detailed formulation of the fourth law in Section \ref{Sec4} and one of its consequences in Section \ref{RCCESEA}.

\begin{figure}[!t]
	\subfloat[Element of a fluid continuum]{{
			\includegraphics[height=0.34\textwidth]{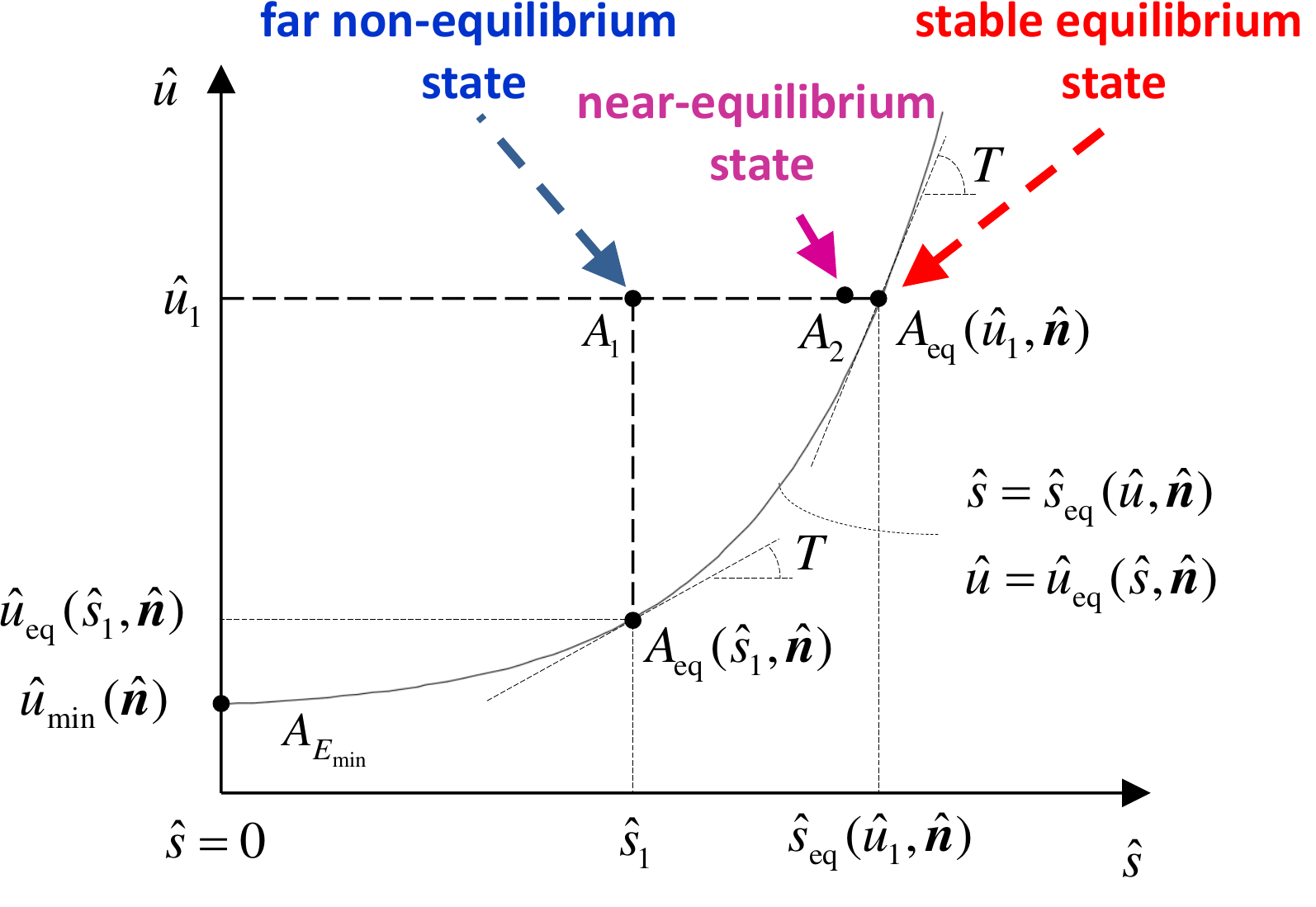}}}
	\subfloat[Closed and uncorrelated quantum system with energy spectrum not bounded from above and ground states non-degenerate]{{
			\includegraphics[height=0.34\textwidth]{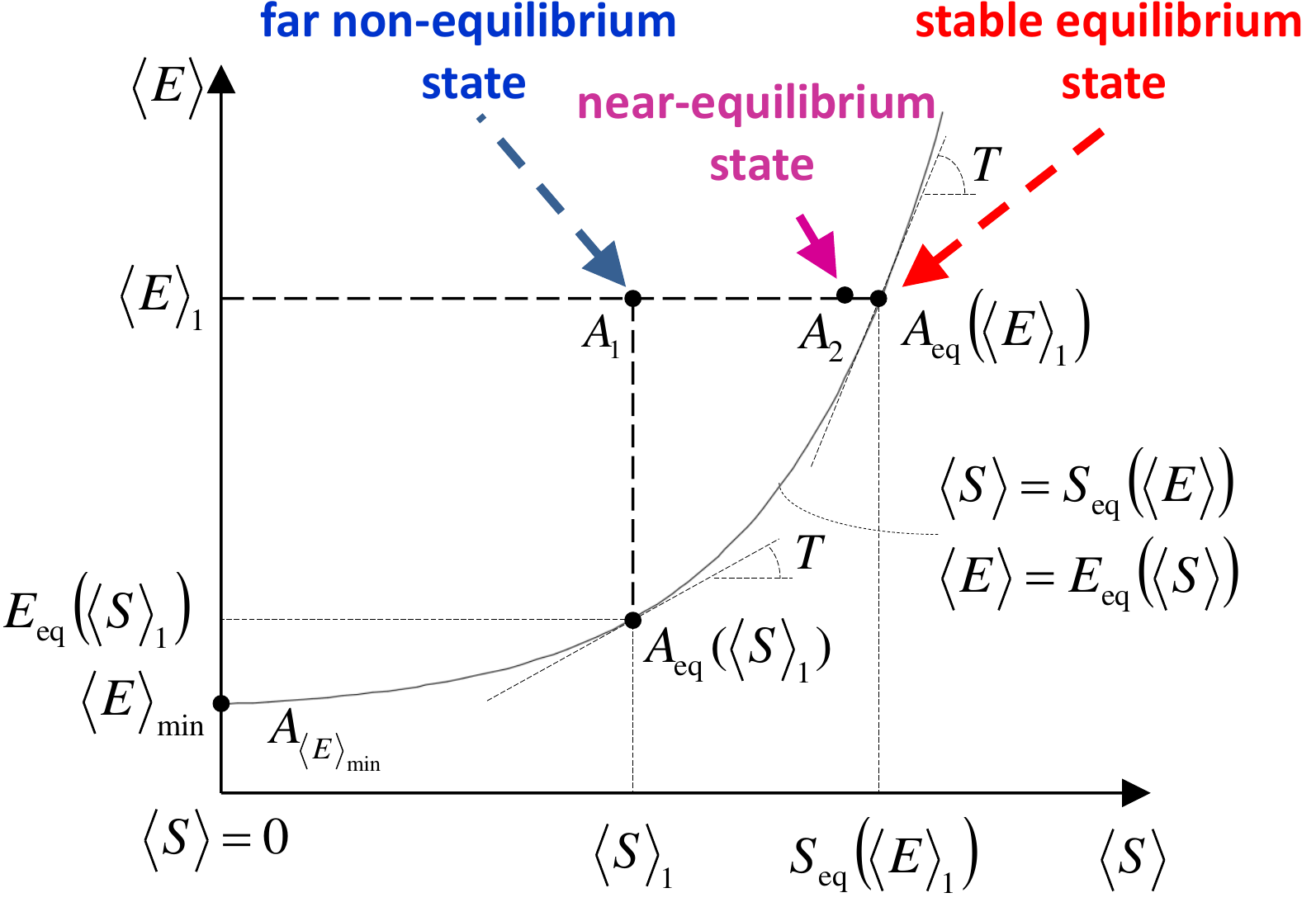}}
	}
	\caption{State representation on the nonequilibrium energy versus entropy diagram \cite{HG3,Book}: (a) for an infinitesimal element of a continuum, $\hat e$, $\hat s$, $\cu$ denote respectively energy, entropy, and amounts of constituents per unit volume, and the fundamental stable-equilibrium relation is $\hat s=\hat s_{\rm eq}(\hat e,\cu)$;   (b) for a closed and uncorrelated quantum system such as a harmonic oscillator, $\langle E\rangle=\Tr(H\rho)$ is the energy, $\langle S\rangle=-\Boltz\Tr(\rho\ln\rho)$ the entropy, and $\langle S\rangle=\langle S\rangle_{\rm eq}(\langle E\rangle)$  the fundamental  stable-equilibrium (Gibbs-state) relation.  }
	\label{Figure2}
\end{figure}

\section{\label{Sec2}Representation on the nonequilibrium energy--entropy diagram} 

The second law implies also the well-known ``maximum entropy principle,'' which states \cite[p.119]{Book}) that among all the states of the system that (within the given model) share the same (mean) value of the energy, the same values of the external control parameters  (if any),   and the same (mean) values of the other independent conserved properties (if any), only the (unique) stable equilibrium state has the maximal entropy. Therefore, for example, respectively, for (a) an infinitesimal element of a fluid continuum with energy density $\hat e$, concentrations $\cu=\{\hat n_1,\dots,\hat n_r\}$, and  entropy density $\hat s$, or (b) a closed and uncorrelated quantum system with mean  energy $\langle E\rangle=\Tr(H\rho)$ and  (nonequilibrium) entropy  $\langle S\rangle=-\Boltz\Tr(\rho\ln\rho)$, the nonequilibrium or non-stable-equilibrium states have entropies strictly smaller than the maximum,
\begin{equation}\label{MaxEnt}
\hat s<\hat s_{\rm eq}(\hat e,\cu)\quad\mbox{case (a)}\qquad\qquad  \langle S\rangle<\langle S\rangle_{\rm eq}(\langle E\rangle)\quad\mbox{case (b)}
\end{equation}

This prompts the energy--entropy diagram representation of nonequilibrium states shown in Figure \ref{Figure2}. It is obtained by first foliating the full state space of the system with respect to the values of its external control parameters and the mean values of the independent conserved properties other than energy, and then by projecting one of these leaves onto the energy--entropy plane. This representation has been  first introduced in \cite{HG3} and fully exploited and explained in \cite{Book}. Recently it has been reintroduced and applied in the quantum thermodynamics framework in \cite{Sparaciari17}.\footnote{This representation is conceptually different from (and must not be confused with)  the representation on the equilibrium energy--entropy diagrams introduced by Gibbs \cite{Gibbs} and used, e.g., in \cite[Par.20]{Landau} and \cite[Fig.1.1]{Mauri}, which refer and are restricted to the equilibrium states  of a system or fluid element in contact with a thermal bath.} 
Temperature is defined only for the stable equilibrium states: (a) $T^{\rm eq}=[\partial \hat s_{\rm eq}(\hat e,\cu)/\partial\hat e]^{-1}$, (b) $T^{\rm eq}=[\partial \langle S\rangle_{\rm eq}(\langle E\rangle)/\partial\langle E\rangle]^{-1}$, and on the energy--entropy diagram it is represented by the slope of the curve representing the fundamental  equilibrium  relation: (a) $\hat s_{\rm eq}(\hat e,\cu)$,   (b) $\langle S\rangle_{\rm eq}(\langle E\rangle)$.

The \caporali{third law of thermodynamics} asserts that the stable equilibrium state of lowest energy [for the given  values of the external control parameters  (if any),   and the given (mean) values of the other independent conserved properties (if any)] has temperature equal to zero and entropy equal to $\Boltz\ln g$ where $g$ is the degeneracy of the corresponding ground state (see 
\cite{Levy12,GBThird15}).

The full description of nonequilibrium states requires in general (in any model) a number of independent variables (typically much) larger than for the equilibrium (maximum entropy) fundamental relation. Denoting by $\pmb\gamma$ the state vector, i.e., the full list of such nonequilibrium independent variables, the entropy and the conserved properties (like all other properties) are functions of such variables
\begin{equation}
\hat s=\hat s({\pmb\gamma})\quad \hat e=\hat e({\pmb\gamma})\quad \cu=\cu({\pmb\gamma})
\quad\mbox{case (a)}\qquad\quad
\langle S\rangle= \langle S\rangle({\pmb\gamma}) \quad \langle E\rangle= \langle E\rangle({\pmb\gamma})  
\quad\mbox{case (b)}
\end{equation}
and the equilibrium fundamental relation and its differential (Gibbs relation) are, respectively,
\begin{equation}
\hat s_{\rm eq}=\hat s({\pmb\gamma}_{{\rm max}\,  \hat s}(\hat e,\cu))
\quad\mbox{case (a)}\qquad\quad
\langle S\rangle_{\rm eq}= \langle S\rangle({\pmb\gamma}_{{\rm max}\,  \langle S\rangle}(\langle E\rangle)) 
\quad\mbox{case (b)}
\end{equation}
\begin{equation}{\rm d} \hat s_{\rm eq}=\beta^{\rm eq}\,{\rm d} \hat e+{\textstyle \sum_i }\lambda_i^{\rm eq}\,{\rm d} \hat n_i\quad\mbox{case (a)}\qquad\qquad\qquad\quad
{\rm d} \langle S\rangle_{\rm eq}=\beta^{\rm eq}\,{\rm d} \langle E\rangle\quad\mbox{case (b)}
\end{equation}
where ${\pmb\gamma}_{\rm eq}={\pmb\gamma}_{{\rm max}\,  \hat s}(\hat e,\cu)$ and ${\pmb\gamma}_{\rm eq}={\pmb\gamma}_{{\rm max}\,  \langle S\rangle}(\langle E\rangle)$ are the solutions of the respective constrained maximum entropy problems: (a) max$_{\pmb\gamma}\ \hat s({\pmb\gamma})$ subject to $\hat e({\pmb\gamma})= \hat e$ and $\cu({\pmb\gamma})=\cu$ for given values of $\hat e$ and $\cu$; and (b) max$_{\pmb\gamma}\ \langle S\rangle({\pmb\gamma})$ subject to $\langle E\rangle({\pmb\gamma})= \langle E\rangle$  for given value of $\langle E\rangle$. Of course, in case (a) $\beta^{\rm eq}=\partial \hat s({\pmb\gamma}_{{\rm max}\,  \hat s}(\hat e,\cu))/\partial \hat e$ and $\lambda_i^{\rm eq}=\partial \hat s({\pmb\gamma}_{{\rm max}\,  \hat s}(\hat e,\cu))/\partial \hat n_i$, and in case (b) $\beta^{\rm eq}=\partial \langle S\rangle({\pmb\gamma}_{{\rm max}\,  \langle S\rangle}(\langle E\rangle))/\partial \langle E\rangle$.

\begin{figure}[!t]
	\begin{center}
	\subfloat[Element of a fluid continuum]{{
			\includegraphics[height=0.34\textwidth]{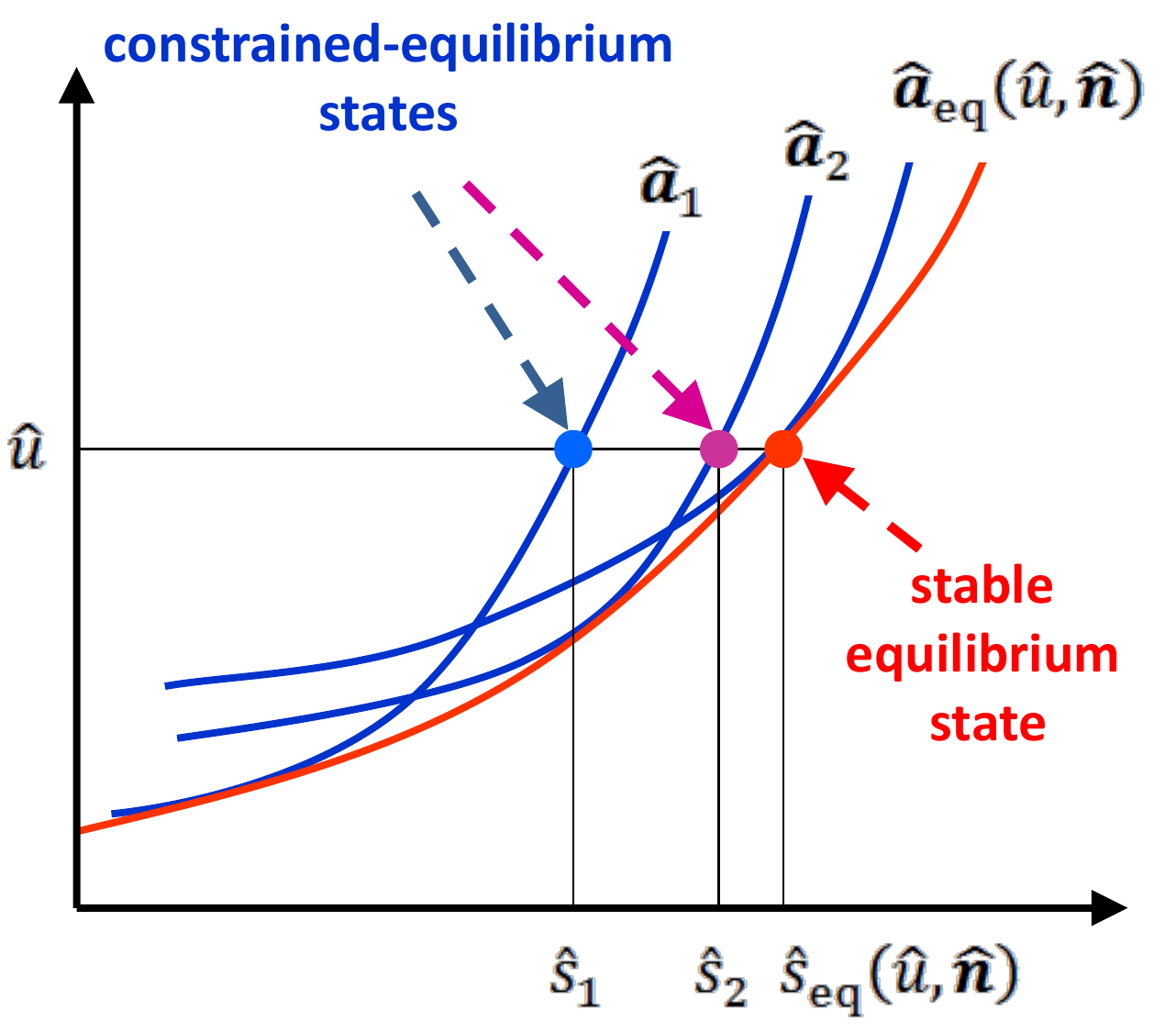}}\hskip5mm}
	\subfloat[Closed and uncorrelated quantum system]{{\hskip5mm
			\includegraphics[height=0.34\textwidth]{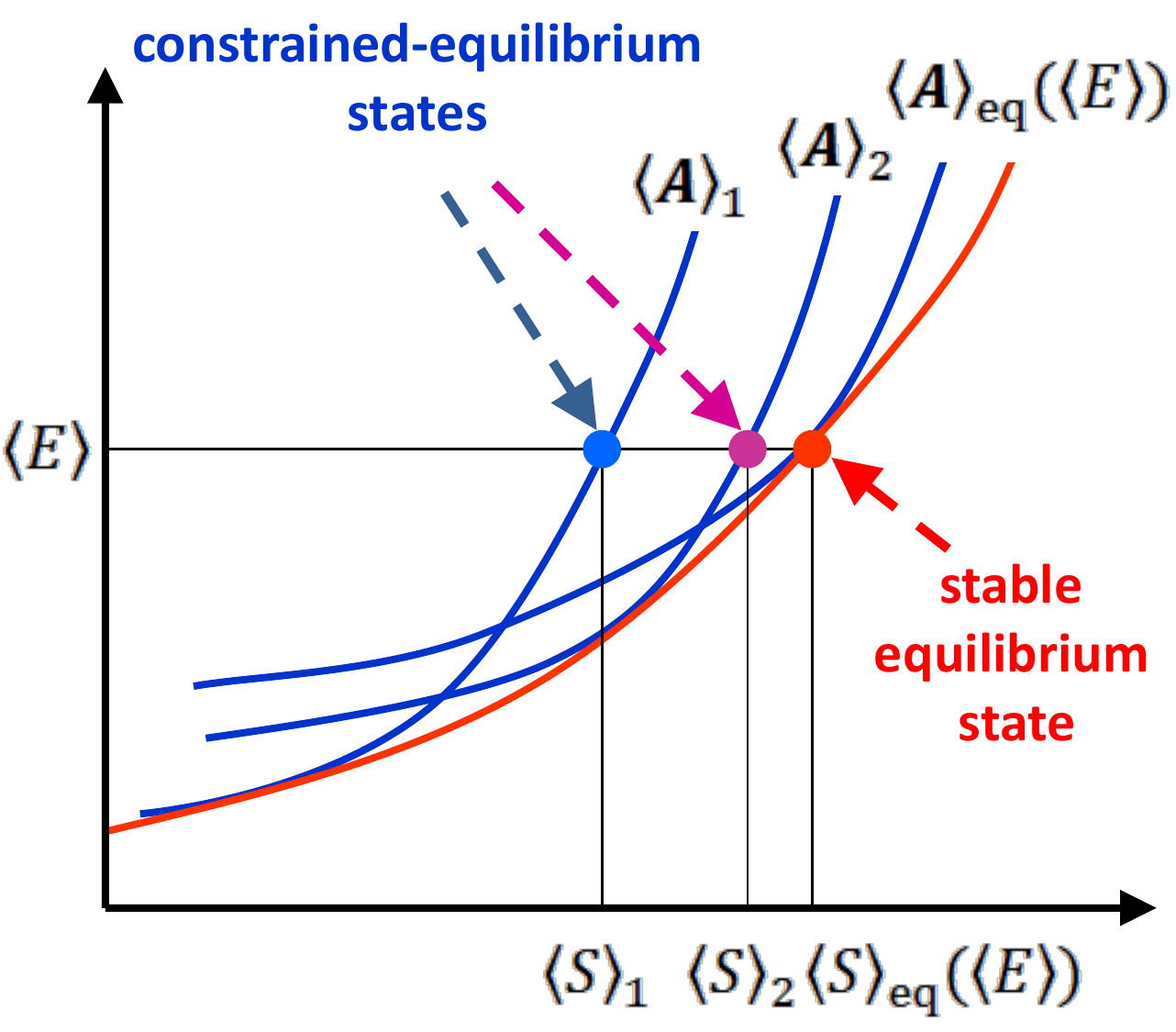}\hskip5mm}
	}\end{center}
	\caption{Representation on the nonequilibrium energy versus entropy diagram of the constrained-equilibrium (quasi-equilibrium) approximation with respect to a set of slow, rate-controlling  state variables: (a) for an infinitesimal element of a continuum, $\hat{ \pmb{a}}=\{\hat{a}_1,\dots,\hat{a}_k,\dots\}$ denotes the set of slowly varying densities;   (b) for a closed and uncorrelated quantum system, $\langle \pmb{A}\rangle=\{\langle  A_1 \rangle,\dots,\langle A_k\rangle,\dots\}$ denotes the set of slowly varying properties $\langle A_k\rangle=\Tr(A_k\rho)$.  }
	\label{Figure3}
\end{figure} 

As part of the ``art'' of choosing the most appropriate level of description, when a detailed description of nonequilibrium states is given in terms of the state variables $\pmb\gamma$ and includes a detailed kinetic law for their time evolution, it  is often possible to identify a small set of slow, rate-controlling (possibly coarse grained) properties, related to the ``bottlenecks'' of the system's detailed kinetics.  We call them the ``rate-controlled constrained-equilibrium'' (RCCE) constraints and denote them by $\hat{ \pmb{a}}({\pmb\gamma})$ in case (a) or $\langle \pmb{A}\rangle({\pmb\gamma})$ in case (b).\footnote{The  ``RCCE approximation''  is a modeling reduction technique introduced and employed extensively by Keck and coworkers \cite{Keck71,Keck90} (see \cite{EntropyRCCE} for key references) in the chemical kinetics and combustion frameworks, where it has inspired a wealth of related and improved model reduction techniques. Recently, prominent authors  (see, e.g., \cite{Gorban01}) have  overlooked the RCCE literature and, by referring to the same method as  ``quasi equilibrium,'' attribute the  idea to an uncited paper in russian \cite{Kogan64}. Unfortunately also the recent \cite{Klika19} fails to discuss relations and differences of their  ``DynMaxEnt'' method   with RCCE. We will show elsewhere that also the recent idea of ``hypo-equilibrium'' \cite{Li2016a} is equivalent to RCCE. Considering that the RCCE method is a ``MaxEnt'' approach, the important connections discussed in \cite{Dewar05,Martyushev13} between maximum entropy production (MEP), fluctuation theorems (FT), minimum entropy production theorems, and maximum dissipation formulations are very much applicable to the RCCE steepest entropy ascent (RCCE--SEA) cases we discuss in Section \ref{RCCESEA}.} The RCCE approximation consists of assuming that the state evolves along the family of maximum entropy manifolds (the blue curves in Figure \ref{Figure3}) parametrized by the  values of the rate-controlling constraints and the conserved properties
\begin{equation}\label{RCCE}
{\pmb\gamma} \approx {\pmb\gamma}_{\rm RCCE}= {\pmb\gamma}_{{\rm max}\,  \hat s}(\hat e,\cu,\hat{ \pmb{a}})
\quad\mbox{case (a)}\qquad\quad
{\pmb\gamma} \approx {\pmb\gamma}_{\rm RCCE}={\pmb\gamma}_{{\rm max}\,  \langle S\rangle}(\langle E\rangle,\langle \pmb{A}\rangle)
\quad\mbox{case (b)}
\end{equation}
where, in terms of  Lagrange multipliers $\beta$, $ \lambda_i$,  $\chi_k$, the RCCE state  ${\pmb\gamma}_{\rm RCCE}$ is the solution of 
\begin{equation}\label{Lagrange}
\frac{\delta \hat s}{\delta {\pmb\gamma}}=\beta\frac{\delta \hat e}{\delta {\pmb\gamma}}+{\textstyle \sum_i }\lambda_i\frac{\delta \hat n_i}{\delta {\pmb\gamma}}+{\textstyle \sum_k } \chi_k\frac{\delta \hat a_k}{\delta {\pmb\gamma}}\ \ \mbox{case (a)}\quad
\frac{\delta \langle S\rangle}{\delta {\pmb\gamma}}=\beta\frac{\delta \langle E\rangle}{\delta {\pmb\gamma}}+{\textstyle \sum_k } \chi_k\frac{\delta \langle A_k\rangle}{\delta {\pmb\gamma}}\ \ \mbox{case (b)}
\end{equation}
As a result, the approximation provides the RCCE fundamental relation and its differential (RCCE Gibbs relation), respectively,
\begin{equation}
\hat s\approx\hat s({\pmb\gamma}_{\rm RCCE}(\hat e,\cu,\hat{ \pmb{a}}))
\quad\mbox{case (a)}\qquad\quad
\langle S\rangle_{\rm eq}\approx \langle S\rangle({\pmb\gamma}_{\rm RCCE}(\langle E\rangle,\langle \pmb{A}\rangle)) 
\quad\mbox{case (b)}
\end{equation} 
\begin{equation}{\rm d} \hat s=\beta\,{\rm d} \hat e+{\textstyle \sum_i }\lambda_i\,{\rm d} \hat n_i+{\textstyle \sum_k } \chi_k\,{\rm d} \hat a_k\ \ \mbox{case (a)}\qquad {\rm d} \langle S\rangle=\beta\,{\rm d} \langle E\rangle+{\textstyle \sum_k } \chi_k\,{\rm d} \langle A_k\rangle\ \ \mbox{case (b)}
\end{equation} 
Where, of course, in case (a) $\beta=\partial \hat s({\pmb\gamma}_{\rm RCCE}(\hat e,\cu,\hat{ \pmb{a}}))/\partial \hat e$, $\lambda_i=\partial \hat s({\pmb\gamma}_{\rm RCCE}(\hat e,\cu,\hat{ \pmb{a}}))/\partial \hat n_i$, and $\chi_k=\partial \hat s({\pmb\gamma}_{\rm RCCE}(\hat e,\cu,\hat{ \pmb{a}}))/\partial \hat a_k$, and in case (b) $\beta=\partial \langle S\rangle({\pmb\gamma}_{\rm RCCE}(\langle E\rangle,\langle \pmb{A}\rangle))/\partial \langle E\rangle$ and $\chi_k=\partial \langle S\rangle({\pmb\gamma}_{\rm RCCE}(\langle E\rangle,\langle \pmb{A}\rangle))/\partial \langle A_k\rangle$.

\section{\label{Sec3}Unified formulation of  basic nonequilibrium dynamical models} 

The explicit dependence of the entropy on the state variables ${\pmb\gamma}$ varies from model to model and in many frameworks it is a characteristic feature of the system. In Ref.\ \cite{PRE14} we have shown that in spite of the differences in state variables,  the essential elements of   five broad frameworks of  nonequilibrium modeling are based on dynamical laws with similar structure, of either of the two forms
\begin{equation}\label{dynamicalLawStructure}
\frac{\partial {\pmb\gamma}}{\partial t}+\nabla\cdot\textbf{J}^o_{\pmb\gamma}  =\pmb{\cal R}_{\pmb\gamma,t} + {{\pmb\Pi}_{\pmb\gamma}}  \quad\mbox{case (a)}\qquad\quad
\frac{{\rm d} {\pmb\gamma}}{{\rm d} t}  =\pmb{\cal R}_{\pmb\gamma,t} + {{\pmb\Pi}_{\pmb\gamma}} \quad\mbox{case (b)}
\end{equation}
where for case (a) the vector field $\textbf{J}^o_{\pmb\gamma}(\pmb{x},t)$ denotes the vector of the fluxes of the components of the state vector field ${\pmb\gamma}$ --- here,  $\textbf{J}^o_{\pmb\gamma}=\textbf{J}_{\pmb\gamma}+{\pmb\gamma}\,\pmb{v}$ where $ \textbf{J}_{\pmb\gamma}$ is the diffusive flux and $\pmb{v}$  the barycentric velocity of the fluid element, $\pmb{v}=\textbf{J}^o_{M}/\hat{m}$ where $\textbf{J}^o_{M}$ is the Lagrangian mass flux and  $\hat{m}$ the fluid's density ---  and  ${{\pmb\Pi}_{\pmb\gamma}}(\pmb{x},t)$ the dissipative contribution responsible for the local entropy production rate; and for both  cases (a) and (b) --- under the assumption that in spite of its interactions with external systems the system remains continuously well-separable (closed) and uncorrelated --- the term  $\pmb{\cal R}_{\pmb\gamma,t}$ includes both internal reversible contributions [such as the effects of chemical reactions in case (a) or of a time-independent Hamiltonian in case (b)]  as well as the effects of interactions (such as models of work and heat interactions, and/or a time-dependent Hamiltonian) that produce exchanges of properties  with external systems in case (b), whereas the term ${{\pmb\Pi}_{\pmb\gamma}}(t)$ accounts for the internal dissipative (entropy generating) contribution.   In either form (a) or (b), the term ${{\pmb\Pi}_{\pmb\gamma}}$ is the only one responsible for  entropy generation and it is incapable of altering the values of the conserved properties. We call ${{\pmb\Pi}_{\pmb\gamma}}$ the ``dissipation component of the dynamics.''

The (Poisson, Hamiltonian) symplectic structure of  the reversible term $\pmb{\cal R}_{\pmb\gamma,t}$ has been the subject of a large number of studies  starting with \cite{Morrison80,Marsden82}. Starting with  \cite{Kaufman84,Morrison84,Grmela84} for classical fields and \cite{Beretta84NATO,Beretta87LN,Beretta85INTJ} for quantum thermodynamics, many have studied the (Riemannian, steepest entropy ascent) metric structure of  the irreversible term ${{\pmb\Pi}_{\pmb\gamma}}$. The resulting combined structure has been given different names depending on the fields of interest and  points of view of the various authors. The main ones are: ``metriplectic structure'' \cite{Morrison86} (see also \cite{Guha07,Materassi16} and references therein),  ``GENERIC'' (general equation for the nonequilibrium reversible-irreversible
coupling \cite{Ottinger97}, see also \cite{Montefusco2015} for an explicit proof of its equivalence with SEA), ``gradient flows,''  ``stochastic gradient flows,'' and particle models, with  ``large deviation principles'' providing strong links between them \cite{Jordan98,Otto01,Mielke11,Duong13,Peletier14,Reina15,Montefusco18,Zimmer18}.

As shown in \cite{PRE14}, the dynamical equation is of type (a) in several frameworks, including: rarefied gas dynamics and small-scale hydrodynamics \cite[Eq.20]{PRE14}, rational extended thermodynamics, macroscopic nonequilibrium thermodynamics, and chemical kinetics \cite[Eq.35]{PRE14}, mesoscopic nonequilibrium thermodynamics, and continuum mechanics with fluctuations \cite[Eq.42]{PRE14}. It is of type (b) in several other frameworks, including: statistical or information-theoretic models of relaxation to equilibrium \cite[Eq.11]{PRE14}, quantum statistical mechanics, quantum information theory, quantum thermodynamics, mesoscopic nonequilibrium quantum thermodynamics, hypo-equilibrium steepest entropy ascent quantum thermodynamics \cite[Eq.59]{PRE14}. 
In all these frameworks, the  balance equations for the entropy take  the  forms
\begin{equation}\label{entropybalance}
\frac{\partial {\hat s}}{\partial t}+\nabla\cdot\textbf{J}^o_{S}  = \Big(\frac{\delta {\hat s}}{\delta {\pmb\gamma}}\Big|{{\pmb{\cal R}}_{\pmb\gamma}}\Big)+ \Big(\frac{\delta {\hat s}}{\delta {\pmb\gamma}}\Big|{{\pmb\Pi}_{\pmb\gamma}}\Big)   \quad\mbox{(a)}\qquad
\frac{{\rm d} \langle S\rangle}{{\rm d} t}  =  \Big(\frac{\delta \langle S\rangle}{\delta {\pmb\gamma}}\Big|{\pmb{\cal R}_{\pmb\gamma}}\Big) +\Big(\frac{\delta \langle S\rangle}{\delta {\pmb\gamma}}\Big|{{\pmb\Pi}_{\pmb\gamma}}\Big) \quad\mbox{(b)}
\end{equation}
where $\textbf{J}^o_{S}=\big(\frac{\delta {\hat s}}{\delta {\pmb\gamma}}\big|\textbf{J}^o_{\pmb\gamma}\big)=\textbf{J}_{S}+\hat{s}\,\pmb{v}$ is the Lagrangian entropy flux and the symbol $\big(\cdot\big|\cdot\big)$ denotes in each framework a suitable inner product on the vector space that contains all the state vectors $\pmb\gamma$, all the functional derivatives $\delta A(\pmb\gamma)/\delta \pmb\gamma$ of the state functionals that represent the properties (such as entropy, energy, and so on), as well as all the vectors ${\pmb\Pi}_{\pmb\gamma}$. Following \cite{Nicole16}, we call ``charge'' (in \cite{Cimento1,Cimento2} we used instead the term ``generator of the motion'') and denote by  $C_i$ any one of the conserved properties,  such as (a) energy and the amounts of constituents, and (b) energy $\langle H\rangle=\Tr(H\rho)=\Tr(H\gamma\gamma^\dagger)$ and total probability  $\langle I\rangle=\Tr(\rho)=\Tr(I\gamma\gamma^\dagger)$. For case (b), following \cite{Cimento1,Gheorghiu1,Gheorghiu2,ROMP}, it is convenient to adopt a state description in terms of a square-root $\gamma$ of the density operator  ($\rho=\gamma\gamma^\dagger$ or $\rho=\gamma^\dagger\gamma$) so as to ensure the nonnegativity condition $\rho\ge 0$.  The balance equation for any charge $C_i$, any RCCE constraint $A_k$, and any other property  take the same forms as in Eq.\ (\ref{entropybalance}), but for charges the orthogonality condition implies
\begin{equation}\label{orthogonality}
 \Big(\frac{\delta \hat{c}_i}{\delta {\pmb\gamma}}\Big|{\pmb\Pi}_{\pmb\gamma}\Big)=0 \quad\mbox{case (a)}\qquad\quad
 \Big(\frac{\delta \langle C_i\rangle}{\delta {\pmb\gamma}}\Big|{\pmb\Pi}_{\pmb\gamma}\Big)=0 \quad\mbox{case (b)}
\end{equation} 
For case (a),  $\textbf{J}^o_{C_i}=\big(\frac{\delta {\hat c}_i}{\delta {\pmb\gamma}}\big|\textbf{J}^o_{\pmb\gamma}\big)=\textbf{J}_{C_i}+\hat{c}_i\,\pmb{v}$ and, for any property $A$, the mass balance equation $\partial \hat{m}/\partial t+\nabla\cdot\textbf{J}^o_{M}  =0$  implies the Reynolds identity $\partial \hat{a}/\partial t+\nabla\cdot\textbf{J}^o_{a}  = \hat{m}\, {\rm D} \hat{a}/{\rm D} t+\nabla\cdot\textbf{J}_{A}$ with $ {\rm D} \hat{a}/{\rm D} t=\partial \hat{a}/\partial t+\hat{a}\,\pmb{v}$. 
Under the local RCCE assumption  (local equilibrium  when all $\chi_k$'s  vanish) and local diffusion-type interaction between adjacent fluid elements $\big(\textbf{J}_{S} = \sum_i\beta_i\,\textbf{J}_{C_i} + \sum_k  \chi_k\,\textbf{J}_{A_k}\big)$, combining the balance equations for  entropy, the charges, the RCCE constraints, and, for case (a),  momentum,  yields the expressions  
\begin{equation}\label{sigma}
{\Pi}_{S} = {\textstyle \sum_i} \textbf{J}_{C_i}\cdot\nabla \beta_i + {\textstyle \sum_k} \textbf{J}_{A_k}\cdot\nabla \chi_k+\Phi + {\textstyle \sum_k} \chi_k {\Pi}_{A_k} \quad\mbox{(a)}\qquad
{\Pi}_{S}  =   {\textstyle \sum_k} \chi_k {\Pi}_{A_k} \quad\mbox{(b)}
\end{equation}
where $\Phi$ is the dissipation function (see \cite{Antanovskii96,Van01,Van08} and references therein for derivations including nonlocal effects), ${\Pi}_{A_k}=\big(\frac{\delta {\hat a_k}}{\delta {\pmb\gamma}}\big|{{\pmb\Pi}_{\pmb\gamma}}\big)$ in case (a) and ${\Pi}_{A_k}=\big(\frac{\delta \langle A_k\rangle}{\delta {\pmb\gamma}}\big|{{\pmb\Pi}_{\pmb\gamma}}\big)$ in case (b) denote the ``dissipative production
rates of the RCCE variables'', and similarly  ${\Pi}_{S}=\big(\frac{\delta {\hat s}}{\delta {\pmb\gamma}}\big|{{\pmb\Pi}_{\pmb\gamma}}\big)$ in case (a) and ${\Pi}_{S}=\big(\frac{\delta \langle S\rangle}{\delta {\pmb\gamma}}\big|{{\pmb\Pi}_{\pmb\gamma}}\big)$ in case (b), denote the (local) entropy production rate.

\section{\label{Sec4}\caporali{Fourth law of thermodynamics}: the dissipative component of evolution is in a direction of steepest  entropy ascent (SEA)} 

We propose to call \caporali{fourth law of thermodynamics} a general modeling rule that captures  a common essential feature of a wide range of models for the dynamical behaviour of systems far from equilibrium and, therefore, encompasses a large body of known experimental evidence. We propose to state it as a ``steepest entropy ascent principle'' as follows: For every state $\pmb{\gamma}$ of a system (close as well as far from equilibrium) the component of the law of time evolution (tangent vector) that is responsible for entropy generation (dissipation) is determined by a local non-degenerate metric operator $G_{\pmb{\gamma}}$ and a local characteristic time $\tau_{\pmb{\gamma}}$. The smooth functionals that define  the charges (conserved properties, generators of the motion) and the entropy on the basis of the first three laws, define the constant-entropy manifolds on each constant-charges leaf  in state space. The metric operator $G_{\pmb{\gamma}}$ defines  the direction of steepest entropy ascent on the constant-charges leaf passing at $\pmb{\gamma}$. The characteristic time $\tau_{\pmb{\gamma}}$ defines the strength of attraction in such direction. Following  in part a suggestion in \cite{Caticha01}, we call $\tau_{\pmb{\gamma}}$ the ``intrinsic dissipation time'' of the system. The general variational formulation of the SEA principle is discussed in \cite{PRE14}.

The metric operator field $G_{\pmb{\gamma}}$ (for shorthand we use the subscript $\pmb{\gamma}$ to denote that it is a function of the state) defines by the usual Riemannian expressions the length of a segment of a one-parameter curve $\pmb{\gamma}(t)$, such as   $\ell(t_2,t_1) =\int_{t_1}^{t_2}\sqrt{(\dot{\pmb{\gamma}}|G_{\pmb{\gamma}}|\dot{\pmb{\gamma}})}\,{\rm d}t$ and $({\rm d}\ell/{\rm d}t)^2= (\dot{\pmb{\gamma}}|G_{\pmb{\gamma}}|\dot{\pmb{\gamma}})$.

As argued in Ref.\ \cite{PRE14} and  discussed below in Section \ref{RCCESEA}, for states near the stable equilibrium manifold the inverse $G_{\pmb{\gamma}}^{-1}$ of the 
metric operator $G_{\pmb{\gamma}}$ is directly related to the Onsager matrix of generalized conductivities.

\begin{figure}[!t]
 	\begin{center}
 		\subfloat{{
 				\includegraphics[width=0.32\textwidth]{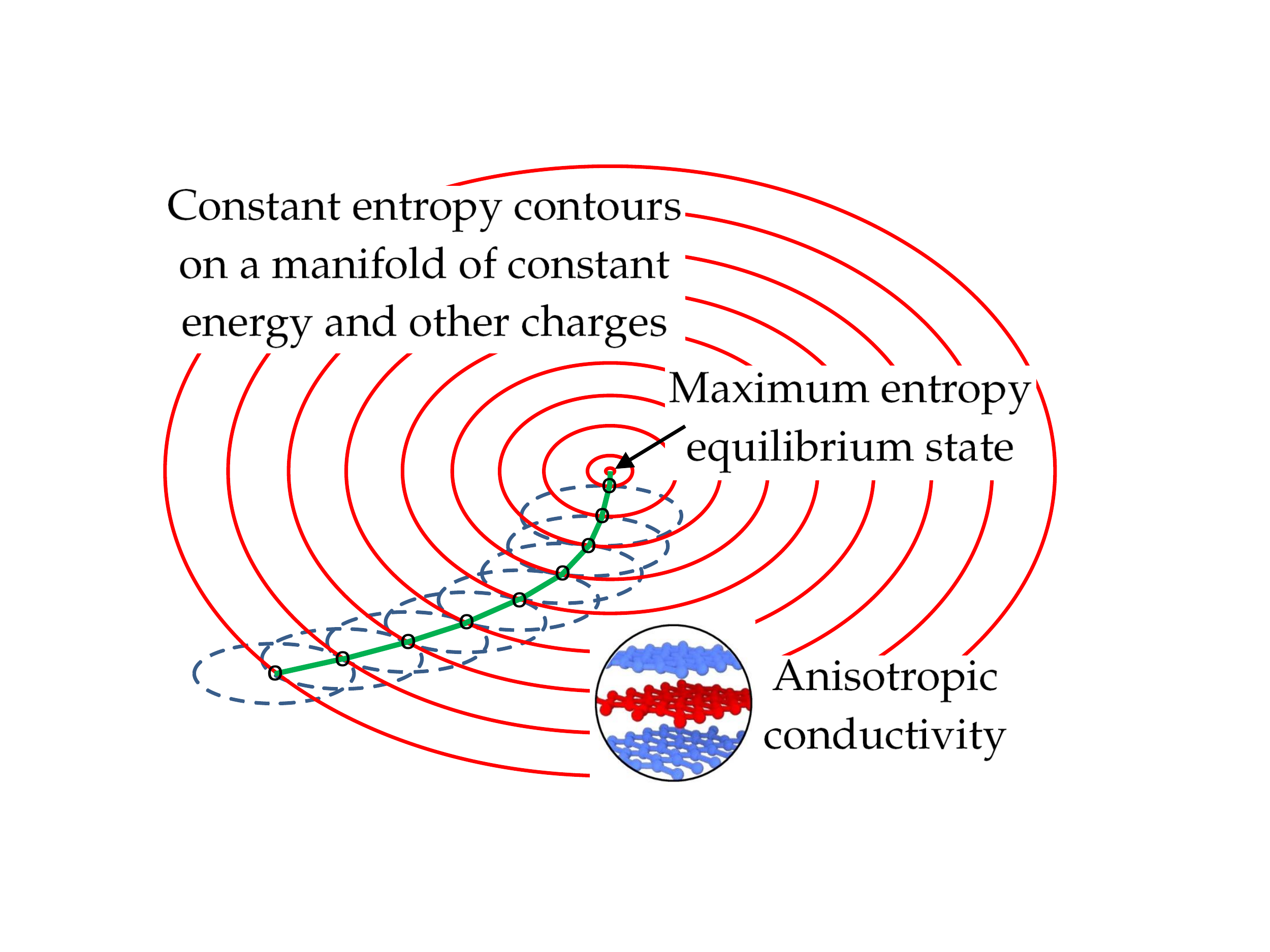}}}
 		\subfloat{{
 				\includegraphics[width=0.32\textwidth]{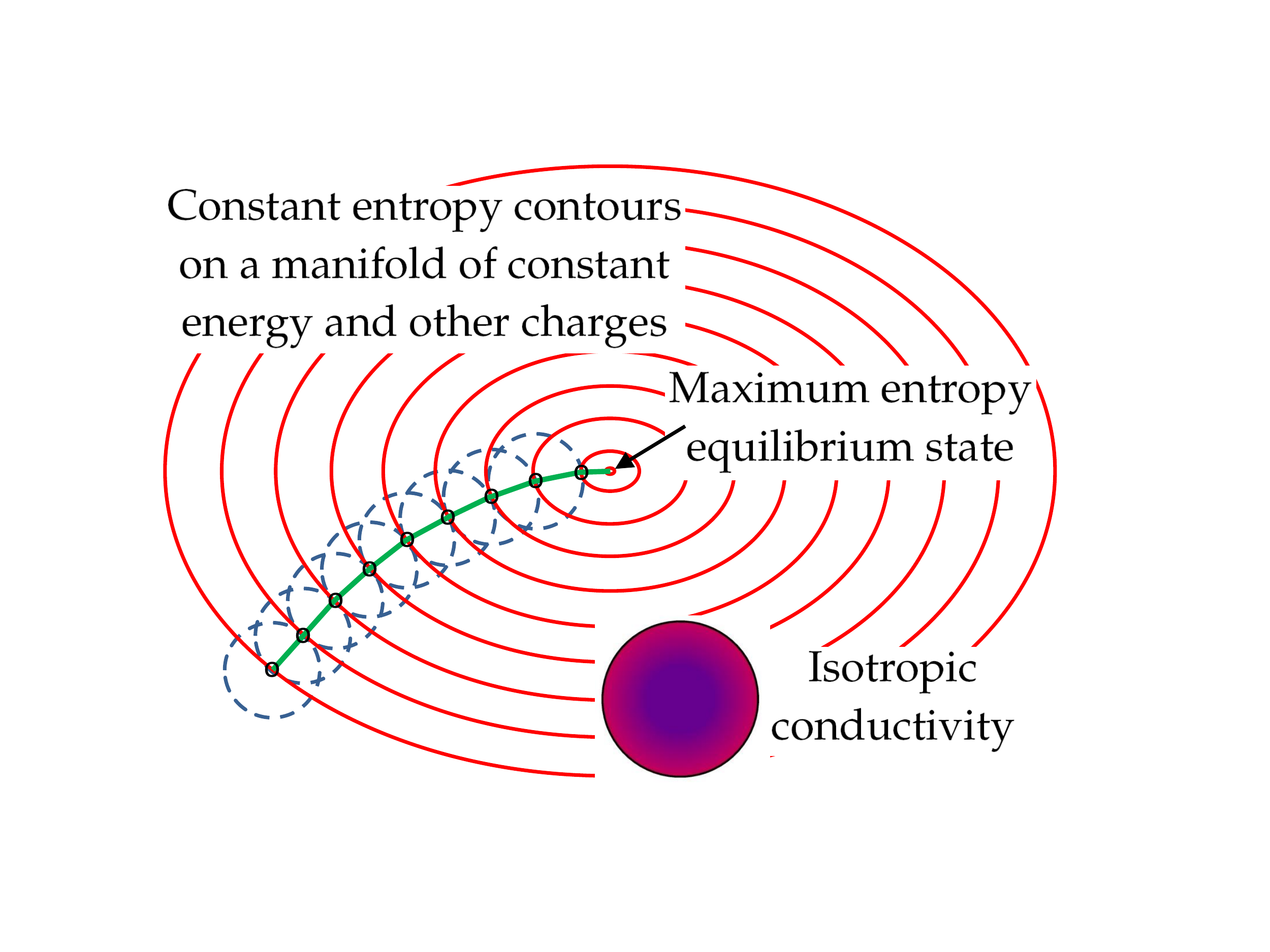}}}
 		\subfloat{{
 				\includegraphics[width=0.32\textwidth]{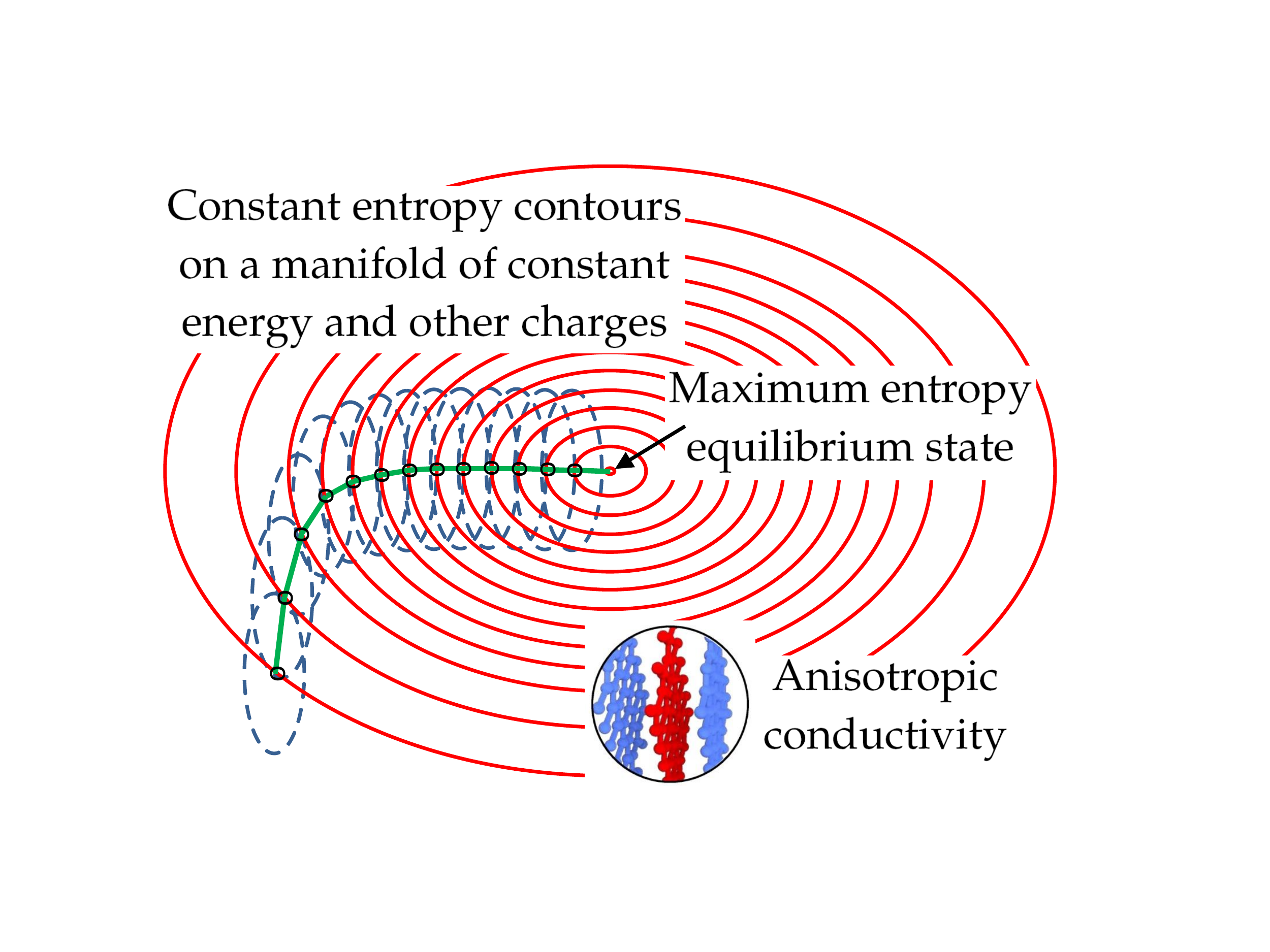}}}
 	\end{center} 
 	\caption{Pictorial representation of steepest entropy ascent evolution for three materials with identical anisotropic entropy landscape (red elliptic contours), identical initial far-nonequilibrium state, but different conductivity tensors (here, for simplicity, assumed state independent): (Left) anisotropic (high horizontal conductivity); (Center) isotropic; (Right) anisotropic (high vertical conductivity). Each blue dashed ellipse (or circle, for the isotropic case) represents a local ball, i.e., the set of  states that (with respect to the local metric) are all at some fixed small distance from the current nonequilibrium state. Among these states, the system chooses to evolve in the direction of the one that has maximal entropy. }
 	\label{Figure4}
 \end{figure}

Two systems $A$ and $B$ with identical kinematics, i.e., identical state
spaces and the same conserved properties, may exhibit different nonequilibrium dynamics, i.e., starting from the same state $\pmb{\gamma}$ they evolve along different paths in state space if they are characterized by different local metric operators 
 $G^A_{\pmb{\gamma}}\ne G^B_{\pmb{\gamma}}$. If instead also the local metric operators are equal, $G^A_{\pmb{\gamma}}= G^B_{\pmb{\gamma}}$, then they evolve along the same path but they may do so at different speeds if the intrinsic dissipation time fields are different, $\tau^A_{\pmb{\gamma}}\ne\tau^B_{\pmb{\gamma}}$.  Figure \ref{Figure4} shows a pictorial representation of SEA evolutions from a far-nonequilibrium state towards stable equilibrium for three  systems with the same anisotropic entropy landscape but different (state independent) metric tensors. To fix ideas this is the case of relaxation to equilibrium of an isolated composite material with  microstructures that yield isotropic or anisotropic thermal conductivity. 
 
The various nonequilibrium modeling approaches and levels of description differ in the bilinear metric forms adopted to define gradients and also in other fine geometrical and mathematical technicalities. The differences between SEA, GENERIC, and metriplectic structures are discussed in \cite[Sec.IIIB]{Montefusco2015}, where we also prove in detail their essential equivalence. The metric that provides a SEA formulation of standard chemical kinetics is available since \cite[Eq.9]{Sieniutycz87}.  The Wasserstein metric operator makes ``gradient flows'' \cite{Jordan98,Otto01,Mielke11}  essentially steepest entropy ascent, when the generating functional is entropy(Lyapunov)-like, i.e., an $S$-function in the sense of \cite{BerettaJMP1986}. The states $\pmb{\gamma}$ are points of a Riemannian manifold ($\cal M$,$ G$) and there is an entropy-like (dimensionless) functional  $\tilde{S}$ on  $\cal M$. In dimensionless time $\tilde{t}=t/\tau_{\pmb{\gamma}}$, the gradient flow of $\tilde{S}$ on ($\cal M$,$ G$) is a dynamical system in $\cal M$ given by the differential equation  ${\rm d}\pmb{\gamma}/{\rm d}\tilde{t}=\mbox{grad} \tilde{S}|_{\pmb{\gamma}}$. The metric operator $G$ is an essential element of the notion. It converts the differential ${\rm diff}\tilde{S}$ of $\tilde{S}$, which is a cotangent vector field, into the gradient of $\tilde{S}$, which is a tangent vector field: for all vector fields $\upsilon$ on ${\cal M}$, $\big({\rm diff}S\big|\upsilon\big) =  \big(\mbox{grad} \tilde{S}\big|G\big| \upsilon\big)  $. Therefore,  for all vector fields $\upsilon$ along $\pmb{\gamma}$, $\big({\rm diff}\tilde{S}|_{\pmb{\gamma}}\big|\upsilon\big) =  \big(\mbox{grad} \tilde{S}|_{\pmb{\gamma}}\big|G_{\pmb{\gamma}}\big| \upsilon\big) =  \big( {\rm d}\pmb{\gamma}/{\rm d}\tilde{t}\big|G_{\pmb{\gamma}}\big| \upsilon\big) $.
The rate of change of the $\tilde{S}$ functional  is
${\rm d}\tilde{S}/{\rm d}\tilde{t}=
\big({\rm diff}\tilde{S}|_{\pmb{\gamma}}\big|{\rm d}\pmb{\gamma}/{\rm d}\tilde{t}\big)=\big({\rm d}\pmb{\gamma}/{\rm d}\tilde{t}\big|G_{\pmb{\gamma}}\big| {\rm d}\pmb{\gamma}/{\rm d}\tilde{t}\big)$, exactly as in the SEA formulation. Moreover, as shown explicitly in \cite{Reina15}, any standard linear diffusion model, where for the diffusive fluxes one assumes  $\textbf{J}_{C_i}=D_{\pmb{\gamma}}^{C_i}\cdot\nabla \beta_i$ and $\textbf{J}_{A_k}=D_{\pmb{\gamma}}^{A_k}\cdot\nabla \chi_k$ in terms of the local diffusion tensors $D_{\pmb{\gamma}}^{C_i}$ and $D_{\pmb{\gamma}}^{A_k}$, are steepest entropy ascent with respect to the (nontrivial) Wasserstein metric operator.

As derived in full details in \cite{ROMP,PRE14,Montefusco2015}, the SEA component of the evolution equation is given by
\begin{equation}\label{SEAEoM}
{\pmb\Pi}_{\pmb\gamma}=\frac{1}{\tau_{\pmb{\gamma}}}G_{\pmb{\gamma}}^{-1}\left(\frac{\delta \hat s}{\delta {\pmb\gamma}}\Big|_C\right)\quad \mbox{case (a)}\qquad\quad {\pmb\Pi}_{\pmb\gamma}=\frac{1}{\tau_{\pmb{\gamma}}}G_{\pmb{\gamma}}^{-1}\left(\frac{\delta \langle S\rangle}{\delta {\pmb\gamma}}\Big|_C\right)\quad \mbox{case (b)}
\end{equation}
where $G_{\pmb\gamma}$ is the local metric operator (it takes and returns vectors on the local constant-charges leaf)  and $\cdot|_C$ denotes  the component of the  variational derivative of the entropy tangent to the local constant-charge manifold, i.e., orthogonal to the variational derivatives of all the charges,
\begin{equation}\label{constrainedgrads}
\frac{\delta \hat s}{\delta {\pmb\gamma}}\Big|_C=\frac{\delta \hat s}{\delta {\pmb\gamma}}-\sum_i \beta_i({\pmb\gamma})\frac{\delta \hat c_i}{\delta {\pmb\gamma}}\quad \mbox{case (a)}\qquad\quad \frac{\delta \langle S\rangle}{\delta {\pmb\gamma}}\Big|_C=
\frac{\delta \langle S\rangle}{\delta {\pmb\gamma}}- \sum_i  \beta_i({\pmb\gamma})\frac{\delta \langle C_i\rangle}{\delta {\pmb\gamma}}\quad \mbox{case (b)}
\end{equation}
and the ``nonequilibrium charge potentials'' $\beta_i({\pmb\gamma})$ are defined at each state $\pmb\gamma$ by the solution of the system of equations expressing such orthogonality conditions, respectively, for the two cases
\begin{equation}\label{potentials}
\Big(\frac{\delta \hat s}{\delta {\pmb\gamma}}\Big|\frac{\delta \hat c_j}{\delta {\pmb\gamma}}\Big)=\sum_i \beta_i({\pmb\gamma})\,\Big(\frac{\delta \hat c_i}{\delta {\pmb\gamma}}\Big|\frac{\delta \hat c_j}{\delta {\pmb\gamma}}\Big)\ \mbox{(a)}\qquad \Big(
\frac{\delta \langle S\rangle}{\delta {\pmb\gamma}}\Big|\frac{\delta \langle C_j\rangle}{\delta {\pmb\gamma}}\Big)= \sum_i  \beta_i({\pmb\gamma})\,\Big(\frac{\delta \langle C_i\rangle}{\delta {\pmb\gamma}}\Big|\frac{\delta \langle C_j\rangle}{\delta {\pmb\gamma}}\Big)\ \mbox{(b)}
\end{equation}

 We have proved in the QT framework \cite{Cimento1,BerettaJMP1986}, and the result
can be readily extended to all other frameworks, that among
the equilibrium states only the maximum entropy one is not
unstable (in the sense of Lyapunov, as specified in \cite{BerettaJMP1986}). As a result, the
maximum entropy states emerge as the only stable equilibrium
ones in the sense of the Hatsopoulos-Keenan statement of the second
law \cite{HK,Book}. In other words, an important part of the (Hatsopoulos-Keenan statement of the) second law emerges as a general theorem of the SEA evolution equation. In addition to meeting all the desiderata formulated in \cite{MPLA1} for strong compatibility with thermodynamics and connecting a variety of important aspects of nonequilibrium, the SEA principle also implies an interesting set of time-energy and time-entropy uncertainty relations \cite{TimeEntropy19}   that allow to  estimate the lifetime of a nonequilibrium state without solving the equation of motion. Moreover, it allows a generalization of Onsager reciprocity to the far nonequilibrium \cite{Beretta87} (the RCCE version is presented below).

Explicit forms of the combined Hamiltonian$+$SEA evolution equation assuming an isotropic (Fisher-Rao) metric ($G_{\pmb{\gamma}}$ the identity operator with $\pmb{\gamma}$  a square root of the density operator) is given in \cite{Beretta85INTJ} for an isolated qubit, in \cite{BerettaIJTP85} for a qubit interacting with a pump-probe laser field, and in \cite{Beretta06} for a four-level qudit. For the isolated qubit  Figure 5 shows the resulting trajectories  inside the Bloch ball, on the $\langle X\rangle$-$\langle Y\rangle$-$S$ constant--$\langle E\rangle$ surface, and on a $\langle E\rangle$-$\langle X\rangle$-$S$ diagram. For applications of the SEA master equation in the framework of quantum computing protocols,  see \cite{Tabakin17}.

\begin{figure}[!t]
	\begin{center}
		\subfloat{{
				\includegraphics[height=0.25\textwidth]{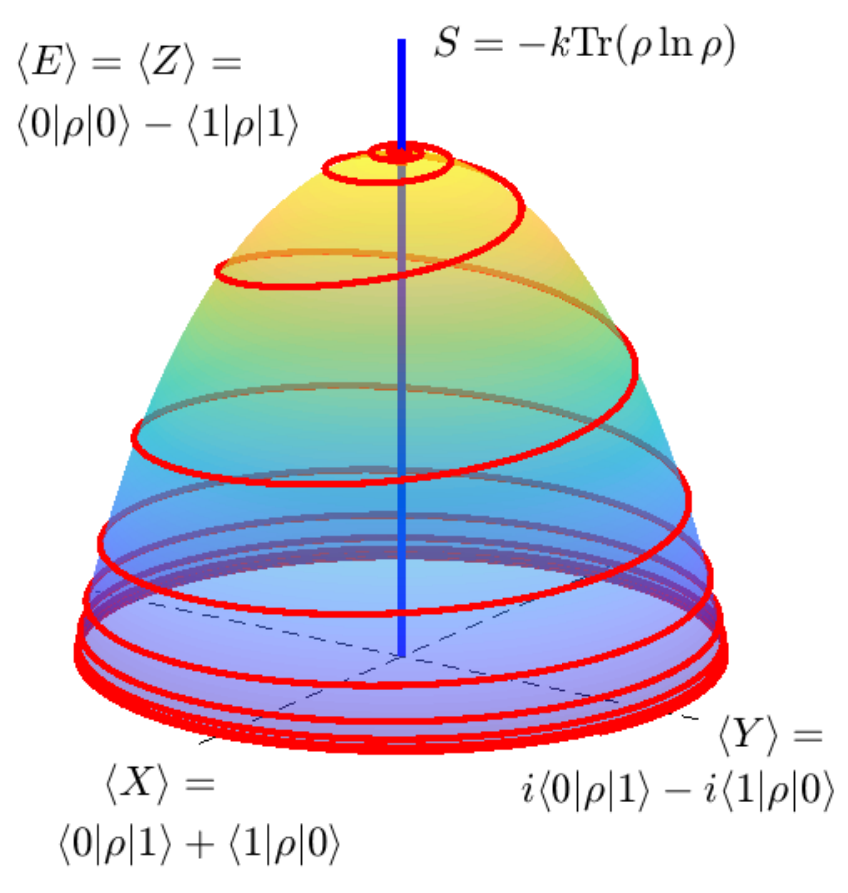}}}
		\subfloat{{
				\includegraphics[height=0.25\textwidth]{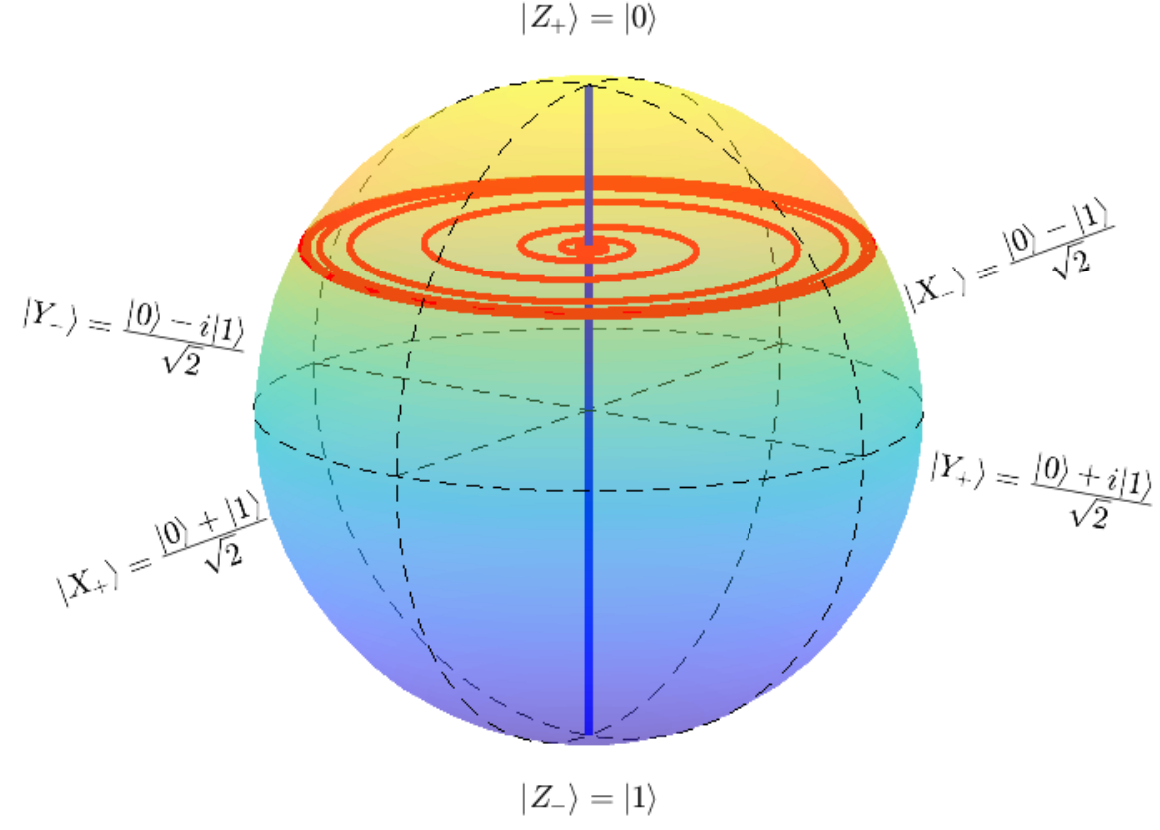}}}
		\subfloat{{
				\includegraphics[height=0.24\textwidth]{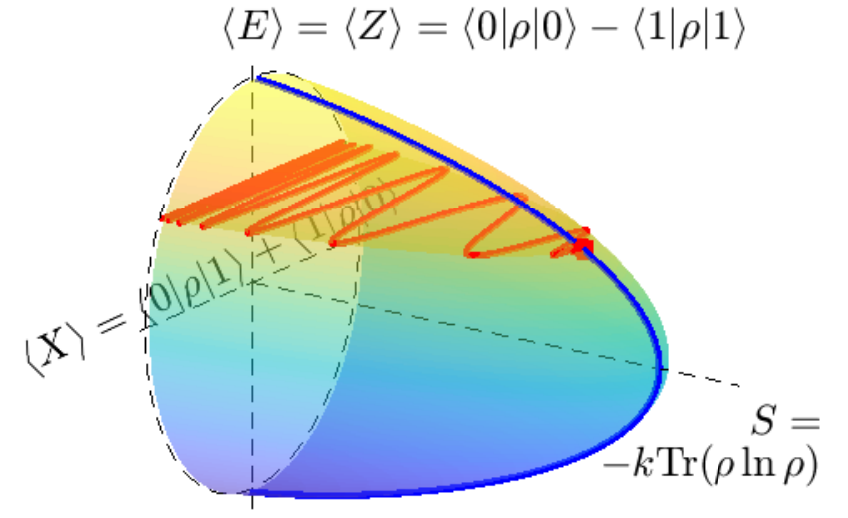}}}
	\end{center} 
	\caption{The states of an isolated qubit map one-to-one with the points of  the Bloch ball: mixed states are  inside, pure states are on the surface (Bloch sphere). A solution of the Hamiltonian$+$SEA(Fisher-Rao) dynamical equation is shown (red curves): (Left) on the $\langle X\rangle$--$\langle Y\rangle$--$S$ constant energy surface;  (Center) inside the Bloch ball; (Right) on the $\langle E\rangle$--$\langle X\rangle$--$S$ diagram. When the trajectory is projected onto the $\langle E\rangle$--$S$ plane, it is a straight constant-energy line approaching asymptotically maximal entropy  for $t\rightarrow\infty$ and zero entropy for  $t\rightarrow-\infty$. As shown in \cite{Beretta85INTJ}, for a state-independent intrinsic dissipation time $\tau$, the rate of entropy production is given by $\frac{{\rm d}S}{{\rm d}t}=\frac{\Boltz}{\tau}\frac{r^2-\langle E\rangle^2}{1-\langle E\rangle^2} \frac{1-r^2}{4r}\big(\ln\frac{1+r}{1-r} \big)^2$ where $r=\sqrt{\langle X\rangle^2+\langle Y\rangle^2+\langle E\rangle^2}$,  $S=-\Boltz\Tr\rho\ln\rho=-\frac{1}{2}\Boltz\left[ (1+r)\ln(1+r)+(1-r)\ln(1-r)\right]$, 
		and energy is relative to a point midway of the two energy levels and scaled by $\hbar\Omega_o$ (where $\Omega_o$ is the Larmor angular frequency), so that $\langle E\rangle=\langle Z\rangle$.}
	\label{Figure5}
\end{figure}

\section{\label{RCCESEA}Far-non-equilibrium RCCE extension of Onsager reciprocity} 

As shown in \cite{Beretta87,PRE14} in the QT framework, for states belonging to a constrained maximal entropy manifold, such as within the RCCE approximation, any SEA evolution equation (i.e., regardless of the particular form  of the dissipative metric operator $G_{\pmb\gamma}$) entails a natural extension of Onsager's reciprocity theorem beyond the near-equilibrium realm. In fact, when Eq.\ (\ref{Lagrange}) holds for the functional derivative of the entropy, Eqs.\ (\ref{constrainedgrads}) and (\ref{potentials}) become, in terms of the ``nonequilibrium constraint potentials'' $\chi_k$ and  the ``projected functional derivatives of the constraints,''
\begin{equation}
\frac{\delta \hat s}{\delta {\pmb\gamma}}\Big|_C= \sum_k \chi_k\frac{\delta \hat a_k}{\delta {\pmb\gamma}}\Big|_C\quad \mbox{case (a)}\qquad\quad \frac{\delta \langle S\rangle}{\delta {\pmb\gamma}}\Big|_C= \sum_k  \chi_k\frac{\delta \langle A_k\rangle}{\delta {\pmb\gamma}}\Big|_C\quad \mbox{case (b)}
\end{equation}
\begin{equation}\label{constrainedgrada}
\frac{\delta \hat a_k}{\delta {\pmb\gamma}}\Big|_C=\frac{\delta \hat a_k}{\delta {\pmb\gamma}}-\sum_i \alpha_{ki}({\pmb\gamma})\frac{\delta \hat c_i}{\delta {\pmb\gamma}}\quad \mbox{(a)}\qquad\quad \frac{\delta \langle A_k\rangle}{\delta {\pmb\gamma}}\Big|_C=
\frac{\delta \langle A_k\rangle}{\delta {\pmb\gamma}}- \sum_i  \alpha_{ki}({\pmb\gamma})\frac{\delta \langle C_i\rangle}{\delta {\pmb\gamma}}\quad \mbox{(b)}
\end{equation}
where the ``partial nonequilibrium constraint potentials'' $\alpha_{ki}({\pmb\gamma})$ are defined by the solution, for each $k$, of the system of equations expressing the othogonality conditions
\begin{equation}
\Big(\frac{\delta \hat a_k}{\delta {\pmb\gamma}}\Big|\frac{\delta \hat c_j}{\delta {\pmb\gamma}}\Big)=\sum_i \alpha_{ki}({\pmb\gamma})\,\Big(\frac{\delta \hat c_i}{\delta {\pmb\gamma}}\Big|\frac{\delta \hat c_j}{\delta {\pmb\gamma}}\Big)\qquad \Big(
\frac{\delta \langle A_k\rangle}{\delta {\pmb\gamma}}\Big|\frac{\delta \langle C_j\rangle}{\delta {\pmb\gamma}}\Big)= \sum_i  \alpha_{ki}({\pmb\gamma})\,\Big(\frac{\delta \langle C_i\rangle}{\delta {\pmb\gamma}}\Big|\frac{\delta \langle C_j\rangle}{\delta {\pmb\gamma}}\Big)
\end{equation}
Thus, finally, by defining the ``RCCE nonequilibrium Onsager generalized conductivities'' 
\begin{equation}\label{Onsager}
L_{jk}({\pmb\gamma})=\frac{1}{\tau_{\pmb{\gamma}}}\left.\left.\left(\frac{\delta \hat a_j}{\delta {\pmb\gamma}}\Big|_C\right| G^{-1}_{\pmb\gamma}\right|\frac{\delta \hat a_k}{\delta {\pmb\gamma}}\Big|_C\right)\qquad 
L_{jk}({\pmb\gamma})=\frac{1}{\tau_{\pmb{\gamma}}}\left.\left.\left(\frac{\delta \langle A_j\rangle}{\delta {\pmb\gamma}}\Big|_C\right| G^{-1}_{\pmb\gamma}\right|\frac{\delta \langle A_k\rangle}{\delta {\pmb\gamma}}\Big|_C\right)
\end{equation}
the SEA component of the evolution equation (\ref{SEAEoM}) and the entropy production can be written as
\begin{equation}\label{SEARCCEEoM}
{\pmb\Pi}_{\pmb\gamma}=\frac{1}{\tau_{\pmb{\gamma}}}\sum_k G_{\pmb{\gamma}}^{-1}\left( \chi_k\frac{\delta \hat a_k}{\delta {\pmb\gamma}}\Big|_C\right)\quad \mbox{(a)}\qquad\quad {\pmb\Pi}_{\pmb\gamma}=\frac{1}{\tau_{\pmb{\gamma}}}\sum_k G_{\pmb{\gamma}}^{-1}\left( \chi_k\frac{\delta \langle A_k\rangle}{\delta {\pmb\gamma}}\Big|_C\right)\quad \mbox{(b)}
\end{equation}
\begin{equation}\label{sdot}
{\Pi}_{S}=\Big(\frac{\delta {\hat s}}{\delta {\pmb\gamma}}\Big|{{\pmb\Pi}_{\pmb\gamma}}\Big)=\sum_j\sum_k\chi_j\, L_{jk}({\pmb\gamma})\,\chi_k\qquad {\Pi}_{S}=\Big(\frac{\delta \langle S\rangle}{\delta {\pmb\gamma}}\Big|{{\pmb\Pi}_{\pmb\gamma}}\Big)=\sum_j\sum_k\chi_j\, L_{jk}({\pmb\gamma})\,\chi_k
\end{equation} 
The natural properties of symmetry and positive definiteness of the non-degenerate metric $G_{\pmb{\gamma}}$ grant automatically (no additional assumptions needed) its invertibility ($G_{\pmb{\gamma}}^{-1}$) and the symmetry and non-negative definiteness of matrix $L_{jk}$. In both cases we have
	\begin{equation}\label{linearNET}
	 {\Pi}_{A_k}=\sum_j \chi_j\,L_{jk}({\pmb\gamma})
	\end{equation}
	where we emphasize that the relations between rates ${\Pi}_{A_k}$ and affinities $\chi_k$ are nonlinear because the $L_{jk}$'s depend on $\pmb\gamma$ which in turn (in the RCCE approximation) are nonlinear functions of the $\chi_k$'s. For a number of references and important comments on the history of internal variables (here, the $\pmb\gamma$'s) and variational formulations of non-equilibrium thermodynamics that  lead to the quasi-linear structure of  Eqs.\ (\ref{sdot}) and (\ref{linearNET}), see \cite{Van08}. Notice that here Eqs.\ (\ref{Onsager}) provide explicit expressions for the $L_{jk}({\pmb\gamma})$'s in terms of the SEA metric $G_{\pmb\gamma}$, the intrinsic dissipation time $\tau_{\pmb\gamma}$, and the projected functional derivatives of the RCCE constraints. Only in the near-equilibrium region we can approximate $L_{jk}({\pmb\gamma})$ with its stable-equilibrium value $L_{jk}({\pmb\gamma}_{\rm eq})$ and obtain the usual nonequilibrium linear rate-affinity relations and near-equilibrium Onsager reciprocity. 
	For space limitations, we cannot pursue this further, but we will show elsewhere that most of the results and discussion presented in \cite{Dewar05} for the case when the entropy is given by $-\sum_i p_i\ln p_i$ can be reformulated also in the present more general context. 
	
	We have shown in \cite{Beretta87,Bregenz09} that Eqs.\ (\ref{Onsager}) which, again, provide explicit relations between the generalized far-nonequilibrium conductivities and the projected functional derivatives of the rate controlling constraints,   have the form of a Gram matrix and represent a far-nonequilibrium generalization of the fluctuation-dissipation theorem.

\section{Conclusion}
Four general rules of thermodynamic modeling reveal four laws of Nature: (1) when the system  is well separated from its environment, its energy must be defined for all states and must emerge as an additive, exchangeable, and conserved property; (2a) when the system is uncorrelated from any other system, its entropy  must be defined  for all states (equilibrium and nonequilibrium) and must emerge as an additive property, exchangeable with other  systems as a result of temporary interactions, conserved in reversible processes and spontaneously generated in irreversible processes; (2b) for  given values of the externally controllable parameters and of the conserved properties other than energy, the states that maximize entropy for a given value of the energy must be the only conditionally locally stable equilibrium points  of the dynamical model (in the sense of \cite[Def.8]{BerettaJMP1986}); (3) among the stable equilibrium states, those with  lowest  energy must have zero temperature; (4) 
every nonequilibrium  state of a system or local subsystem for which entropy is well-defined must be equipped with a metric in state space  with respect to which the irreversible component of its time evolution is in the direction of steepest entropy ascent compatible with the conservation constraints.

Rules (1) to (3) are  well-known essential features/consequences of the first, second, and third law of thermodynamics, respectively. 
Our main point in this paper is that an enormous body of scientific research  devoted to modeling the essential features of nonequilibrium natural phenomena during the past four decades has converged from many different directions and frameworks towards the general recognition (albeit still expressed in different but equivalent forms and language) that also Rule (4) is  indispensable. For this reason, we claim that it reveals a great law of Nature and, therefore, we propose to call it the fourth law of thermodynamics.

To illustrate the power of the proposed fourth law, we provide in Section \ref{RCCESEA} a new proof that, within the framework of the rate-controlled constrained-equilibrium (RCCE) approximation (also known as the quasi-equilibrium approximation), it allows to extend Onsager reciprocity and  fluctuation-dissipation relations --- which are well-known signatures of nonequilibrium dynamics in the near-equilibrium neighborhood --- to the entire far-nonequilibrium state space, where the relations between affinities and dissipative rates (force-flux relations) are nonlinear, but have a quasi-linear structure.

The impressive revival of interest on thermodynamics over the past two decades has been fueled by the increasing roles that  thermodynamics and quantum thermodynamics have started playing in a wide range of emerging and prospective technologies.  Studies in these fields have evolved quite independently, and, for a long while, researchers from different fields (mechanical engineering, continuum mechanics, solid mechanics, physics, chemical engineering, nonequilibrium thermodynamics, quantum thermodynamics, mathematical physics) have developed their  ideas often unaware of parallel developments ongoing or already done in other fields. Efforts like the present one to connect,  distill, merge, and  unify  the essentials of these sparse contributions have already started, but it will take several years to fill completely the gap.

Many will argue that in some nonequilibrium frameworks steepest entropy ascent  is an invalid or  unnecessary principle. For example, a Referee insisted on the following remark (inserted here per explicit request of the Editor):  \caporali{The ``Steepest Entropy Ascent'' may not be valid in Stochastic Thermodynamics where processes of negative entropy production exist. The conventional Non-Equilibrium Thermodynamics consisting of state space, balance equations, constitutive equations and Second Law, resulting in a system of differential equations solvable by taking constraints into account, does not need a steepest entropy ascent. May be that such a concept is hidden in the conventional procedure, but Stochastic Thermodynamics may be a counter-example.}  

In anticipation of discussions about  the above remark, it is useful to keep  in mind that: (1) the concept of   ``processes with negative entropy production'' (see, e.g., \cite{Bhattacharya17}) has been already criticized (see, e.g., \cite{Esposito10}); (2) in stochastic thermodynamic models of  effects of strong system-bath correlations (such as echoes, recurrences, purity revivals), the microscopic definitions of internal energy, entropy, work, heat, free energy, available energy with respect to a thermal environment, adiabatic availability, \textit{etc.} must satisfy strict consistency conditions  (see, e.g., \cite{Jarzynski17});  and (3)  in such processes, the  irreversible component of dynamics (potentially subject to the fourth law)  is only the part of the evolution equation which is responsible for (fluctuating, but on average progressively) incomplete recurrences (see, e.g., \cite[Fig.6]{Bhattacharya17} and \cite[Fig.1]{Esposito10}), due to  lost or inaccessible correlations: for example, in kinetic theory, the collision integral in the highest order  level of a truncated Bogoliubov-Born-Green-Kirkwood-Yvon (so-called BBGKY) hierarchy.

As  John Maddox (perhaps the most famous editor of \textit{Nature}) wrote 35 year ago  (ten years before becoming an honorary fellow of the Royal Society) in an editorial about one of the earliest attempts to build a quantum thermodynamics \cite{Maddox85}, ``this is a field in which the proof of the pudding is in the eating.''

\vskip6pt

\enlargethispage{20pt}









\end{document}